\documentclass[12pt]{article}
\def\gappeq{\mathrel{\rlap {\raise.5ex\hbox{$>$}}
{\lower.5ex\hbox{$\sim$}}}}

\def\lappeq{\mathrel{\rlap{\raise.5ex\hbox{$<$}}
{\lower.5ex\hbox{$\sim$}}}}

\def\Toprel#1\over#2{\mathrel{\mathop{#2}\limits^{#1}}}

\begin{document}
\pagestyle{empty}
\begin{flushright}
{CERN-TH/2000-343}\\
hep-ph/0012014
\end{flushright}
\vspace*{5mm}
\begin{center}
{\Large \bf Application of new multiloop QCD input\\ 
to the analysis of $xF_3$ data}
 \\
\vspace*{1cm} 
{\bf A. L. Kataev}$^{(a)}$, {\bf G. Parente}$^{(b,1)}$ and {\bf A.V. Sidorov}$
^{(c,2)}$ \\
\vspace{0.3cm}
(a) Theoretical Physics Division, CERN 
CH - 1211 Geneva 23 and  \\
Institute for Nuclear Research of the Academy of Sciences 
of Russia, 117312 Moscow, Russia \\
(b) Department of Particle Physics, University of Santiago de Compostela,\\
15706 Santiago de Compostela, Spain \\
(c) Bogoliubov Laboratory of Theoretical Physics, Joint Institute 
for Nuclear Research, 141980 Dubna, Russia
\end{center}
\vspace*{2cm}  
\begin{center}
{\bf ABSTRACT} 
\end{center}
\vspace*{5mm}
\noindent
The new theoretical input to the analysis of the experimental 
data of the CCFR collaboration for $F_3$ structure function 
of $\nu N$ deep inelastic scattering is considered. This input comes 
from the next-to-next-to-leading order corrections 
to the anomalous dimensions of the Mellin moments of the $F_3$ 
structure function. The QCD scale $\Lambda_{\overline{MS}}^{(4)}$ 
is extracted from higher-twist independent fits. The results 
obtained demonstrate the minimization of the influence of 
perturbative QCD contributions to the value of $\Lambda_{\overline{MS}}^{(4)}$.
\vspace*{0.5cm} 
\noindent 

{\it Based on  Contributed to the Proceedings of Quarks-2000 
International Seminar,
Pushkin, May 2000, Russia and  of ACAT'2000 Workshop, 
Fermilab, October 2000, USA}
\vspace{0.5cm}

\noindent
$^{1}$ Supported by Xunta de Galicia (PGIDT00PX20615PR) and CICYT 
(AEN99-0589-C02-02)\\
$^{2}$ Supported by RFBI (Grants N 99-01-00091, 00-02-17432) 
and by INTAS call 2000 (project N587)
\vspace*{0.5cm}

\begin{flushleft} CERN-TH/2000-343 \\
November  2000
\end{flushleft}
\vfill\eject

\setcounter{page}{1}
\pagestyle{plain}

\begin{center}
{\Large \bf Application of new  multiloop  QCD input\\ to the analysis 
of $xF_3$ data }

{\bf A.L. Kataev$^{a}$, G. Parente$^{b}$ and  
A.V. Sidorov$^{c}$}

{$^{a}$Theoretical Physics  Division,  CERN, CH-1211  Geneva, Switzerland and\\
 Institute for Nuclear Research of the Academy of Sciences 
of Rusia,\\ 117312 Moscow, Russia\\
$^{b}$Department of Particle Physics, 
University of Santiago de Compostela,\\ 15706 Santiago de 
Compostela, Spain\\
$^{c}$ Bogoliubov Laboratory of Theoretical Physics, Joint Institute for Nuclear Research,\\ 141980 Dubna, Russia}

\end{center}

\begin{abstract}
  The new  theoretical  input to the 
analysis of the experimental data  of the CCFR collaboration 
for $F_3$ structure function of $\nu N$ deep inelastic 
scattering  is considered. 
This input comes  from the 
next-to-next-to-leading order corrections to the 
anomalous dimensions of the Mellin moments of the  
$F_3$ structure function and N$^3$LO corrections to the 
related coefficient funtions. 
The  QCD scale parameter 
$\Lambda_{\overline{MS}}^{(4)}$ is extracted from  higher-twist independent 
fits. The results obtained demonstrate the minimization of the influence  
of  perturbative QCD contributions to the  value 
of $\Lambda_{\overline{MS}}^{(4)}$.  
\end{abstract}



\section{Introduction}

One of the most important current problems of symbolic perturbative 
QCD studies is the analytical evaluation of the next-to-next-to-leading 
order (NNLO) QCD corrections to the kernels of the DGLAP equations
~\cite{DGLAP} for different structure functions of the 
deep-inelastic scattering 
(DIS) process. In this  note we will apply 
the related information for the  
fixation of definite uncertainties of the NNLO 
analysis~\cite{KKPS,KPS1} of experimental 
data for $F_3$ structure function (SF) data of $\nu N$ DIS, provided by the 
CCFR collaboration~\cite{CCFR} at the
Fermilab Tevatron and present preliminary 
results of our improved fits which  will  be described elsewhere~\cite{KPS2}.

\section{Methods of analysis of DIS data}

There are several methods of  analysis of the experimental data 
of DIS in the high orders of perturbation theory. The traditional 
method is based on the solution of the DGLAP equation, which in the  case 
of the $F_3$ SF has the following form:
\begin{equation}
Q^2\frac{d}{dQ^2}F_3(x,Q^2)=\frac{1}{2}\int_x^{1}\frac{dy}{y}
\bigg[V_{F_3}(y,A_s)+\beta(A_s)\frac{\partial{\rm ln}C_{F_3}(y,A_s)}
{\partial A_s}\bigg]F_3\bigg(\frac{x}{y},Q^2\bigg)
\end{equation}
where $A_s=\alpha_s/(4\pi)$, $\mu\partial A_s/\partial\mu=\beta(A_s)$ 
is the QCD $\beta$-function and  $C_{F_3}(y,A_s)$ is the coefficient function, defined as   
\begin{equation}
C_{F_3}(y,A_s)=\sum_{n\geq 0} C_{F_3,n}(y)
\bigg(\frac{\alpha_s}{4\pi}\bigg)^{n}
\end{equation}
and $V_{F_3}(z)$ is the DGLAP kernel, related to a non-singlet (NS) 
$F_3$ SF. The solution of Eq.(1) is describing the 
predicted by perturbative QCD violation of scaling 
\cite{Bj} or automedeling \cite{BVT} behaviour of the DIS SFs 
by the logarithmically decreasing order $\alpha_s$-corrections.

The coefficient function we are interested in has been  
known  at the NNLO for  quite a long period.  
The term $C_{F_3,2}(y)$ was analytically calculated in Ref.\cite{VZ}. 
The results 
of these calculations were confirmed recently \cite{MV} using a 
different technique. 

The kernel $V_{F_3}(z,\alpha_s)$ is analytically known only at the 
NLO. However, since there exists a   method of symbolic evaluation 
of multiloop corrections to the renormalization group functions in the 
$\overline{MS}$-scheme \cite{T}
and its realization  at the FORM  system, it became 
possible to calculate analytically the NNLO corrections to the 
$n=2,4,6,8,10$ Mellin moments of the NS kernel of the $F_2$ SF 
\cite{Larin}. They have the following expansion:
\begin{equation}
-\int_0^{1} z^{n-1}V_{NS,F_2}(z,\alpha_s)dz
= \sum_{i\geq 0}\gamma_{NS,F_2}^{(i)}(n)
\bigg(\frac{\alpha_s}{4\pi}\bigg)^{i+1}
\end{equation}
and are related 
to the anomalous dimension of 
NS renormalization group (RG) constants of $F_2$ SF\footnote
{The method of renormalization group was originally developed in \cite{RG}.} : 
\begin{equation}
\mu\frac{\partial\ln Z_n^{NS,F_2}}{\partial\mu}
=\gamma_{NS,F_2}^{(n)}(\alpha_s)~~~~.
\end{equation}

These results were used in the process of the fits of Refs.\cite{KKPS,KPS1} 
of the  CCFR data for the $F_3$ SF  
with the help of the Jacobi polynomial method \cite{Jacobi}. It allows 
the reconstruction of  the SF $F_3$ from the {\bf finite} number of 
Mellin moments 
$M_{j,F_3}(Q^2)$ of the  $xF_3$ SF:
\begin{equation}
F_3^{N_{max}}(x,Q^2)=w\sum_{n=0}^{N_{max}}
\Theta_n^{\alpha,\beta}(x)\sum_{j=0}^{n}c_j^{(n)}(\alpha,\beta)
M_{j+2,F_3}^{TMC}(Q^2)   
\end{equation}
where $w=w(\alpha,\beta)=x^{\alpha-1}(1-x)^{\beta}$, $\Theta_n^{\alpha,\beta}$
are the orthogonal Jacobi polynomials and $c_j^{(n)}(\alpha,\beta)$ 
is the combination of Euler $\Gamma$-functions, which is factorially 
increasing with increasing  of  $N_{max}$ and thus $n$.

The expressions for $M_{j+2,F_3}^{TMC}(Q^2)$ include 
the information about Mellin moments of the coefficient function 
\begin{equation}
C_{n,F_3}(Q^2)=\int_0^{1}x^{n-1}C_{F_3}(x,\alpha_s)dx
=\sum_{i\geq 0}C^{(i)}(n)\bigg(\frac{\alpha_s}{4\pi}\bigg)^{i}
\end{equation}
where $C^{(0)}(n)=1$. The target mass corrections, 
proportional to 
 $(M_N^2/Q^2)M_{j+4,F_3}(Q^2)$, are also included 
into the fits.
Therefore, the number of the Jacobi polynomials $N_{max}=6$ 
corresponds to taking into account the information about RG  
evolution of  10 moments, and $N_{max}=9$ presumes that 
the evolution of $n=13$ number of Mellin moments is considered.    

The procedure of reconstruction of $F_3(x,Q^2)$ from the finite 
number of  Mellin moments and the related fits of the experimental 
data were implemented in the form of   FORTRAN programs. 
The details of the fits 
of the CCFR data, based on   RG evolution of 10 moments, 
are desribed in Refs.\cite{KKPS,KPS1} (for the brief review 
see Ref.\cite{KPSB}). 
In the process of these  analyses the
following approximations were made:
a)  it was assumed that for a large enough number of moments, 
$\gamma_{NS,F_3}^{(n)}(\alpha_s)\approx\gamma_{NS,F_2}^{(n)}(\alpha_s)$;
 b)  since the odd NNLO terms of $\gamma_{NS,F_2}^{(n)}$ are explicitly 
unknown, they were fixed using the smooth interpolation procedure proposed 
in Ref.\cite{PKK}.
It was known that  the additional contributions, proportional 
to the $d^{abc}d^{abc}$ structure of the colour gauge group 
$SU(N_c)$  are 
starting to contribute to the coefficients of 
$\gamma_{NS,F_3}^{(n)}(\alpha_s)$ from the NNLO \cite{KPS1}. 
In the process of the 
analysis of Refs.\cite{KKPS,KPS1} it was assumed that they were 
not dominating and therefore were not taken into account. 

\section{New inputs for the fits}

After recent explicit analytical evaluation of 
the NNLO coefficients of $\gamma_{NS,F_3}^{(n)}(\alpha_s)$ at 
$n=$3,5,7\\,9,11,13 (see Ref.\cite{RV}) it became possible to fix 
this uncertainty
(it is worth  noting that  the 
NNLO contribution to $\gamma_{NS,F_2}^{(n)}(\alpha_s)$ for $n=$12 
was  analytically evaluated in Ref.\cite{RV} also). To estimate  
the NNLO terms of $\gamma_{NS,F_3}^{(n)}(\alpha_s)$ at $n=$4,6,8,10,12 
we applied the smooth interpolation procedure, identical to the one 
used to estimate the odd NNLO terms of 
$\gamma_{NS,F_2}^{(n)}(\alpha_s)$, while the numerical value of 
$\gamma_{NS,F_3}^{(2)}(2)$ was fixed  with the help of an extrapolation 
procedure, where we have not used the value at $n=1$. 
The justification and more details of this procedure 
will be given elsewhere \cite{KPS2}.

The used numerical results of the NNLO contributions 
$\gamma_{NS,F_3}^{(2)}(n)$ with and without $d^{abc}d^{abc}$-factors are 
presented in Table 1, where we marked in parenthesis the estimated even 
terms. The expressions for the NNLO
contributions to the NS anomalous dimensions terms $\gamma_{NS,F_2}^{(2)}(n)$
are also given for  comparison. 
They include the numerical results of the explicit 
analytical calculations of Refs.\cite{Larin,RV}, normalized to $f=4$ numbers 
of active flavours, and the results of the smooth 
interpolation procedure, in parenthesis,  
applied for estimating explicitly uncalculated 
odd terms. The satisfactory agreement between the numbers in the second and 
third columns  supports the assumptions a) and b) mentioned above. 
\begin{table}
\begin{tabular}{||r|c|c|c||}
\hline
  $n$ & $\gamma_{NS,F_3}^{(2)}(n)$ &
$d^{abc}d^{abc}$ neglected in  $\gamma_{NS,F_3}^{(2)}(n)$ &
 $\gamma_{NS,F_2}^{(2)}(n)$ \\
\hline
2  & (631)  & (585) & 612.06 \\ 
3  &  861.65   &  836.34 & (838.93)\\  
4  &  (1015.37) & (1001.42) & 1005.82\\  
5  &  1140.90   & 1132.73   & (1135.28)\\
6  &  (1247) & (1241.21) & 1242.01 \\
7  &  1338.27   & 1334.32   & (1334.65) \\
8  &  (1420) & (1416.73) & 1417.45 \\
9  &  1493.47   & 1491.13   & (1492.02)\\
10 &  (1561) & (1558.85) & 1559.01 \\
11 &  1622.28   & 1620.73   & (1619.83)\\            
12 &  (1679.81) & (1677.70) & 1678.40 \\       
\hline
\end{tabular}
\caption{The numerical expressions of the  NNLO coefficients 
of anomalous dimensions of the $n$-th 
NS moments of the $F_3$ and $F_2$ SFs 
at $f=4$. The numbers in parenthesis are the estimated  
results. }
\label{tab:a}
\end{table}
\begin{table}
\begin{tabular}{||r|c|c|c|c||}
\hline
 $n$
 &$C^{(1)}(n)$
 &$C^{(2)}(n)$
 &$C^{(3)}(n)$
 &$C^{(3)}(n)_{[1/1]}$\\
\hline
1  & $-$4    & $-$52  & $-$644.35 & $-$676    \\ 
2  & $-$1.78 & $-$47.47 & ($-$1127.45)   &  $-$1268 \\  
3  &  1.67 & $-$12.72 & $-$1013.17 &   97\\  
4  & 4.87    & 37.12  & ($-$410.66) & 283 \\
5  &  7.75 & 95.41 & 584.94 & 1175  \\
6  &  10.35   & 158.29   & (1893.58)& 2421  \\
7  & 12.72   & 223.90  & 3450.47 & 3940  \\
8  & 14.90    & 290.88 & (5205.39)   & 5679 \\
9 & 16.92 & 358.59 & 7120.99  & 7602  \\
10 & 18.79 & 426.44 & (9170.21) & 9677 \\              
11 & 20.55  & 494.19 & 11332.82  & 11884  \\       
12 & 22.20 & 561.56 & (13590.97) & 14205  \\
13 & 22.76 & 628.45 & 15923.91  & 17353   \\
\hline
\end{tabular}
\caption{The numerical expressions for the coefficients  of the coefficient 
functions for $n$-th Mellin moments of the $F_3$ SF up to N$^3$LO and their 
[1/1] Pad\'e estimates. }
\label{tab:b}
\end{table}



In Table 2  the numerical expressions for the coefficients
of Eq.(6) for $f=4$ numbers of active flavours are given. They  
include the results 
of explicit calculations of N$^3$LO corrections of odd moments \cite{RV},
supplemented with the information about the coefficients of the 
Gross--Llewellyn Smith sum rule \cite{GL,LV},  defined by the  
$n=1$ Mellin moment of the $xF_3$ SF. 
The numbers  in parenthesis are the results 
of the interpolation procedure. In the last column we present the 
values of $C^{(3)}(n)$, obtained with the help of the [1/1] 
Pad\'e estimates approach. One can see that the agreement of 
Pad\'e estimates with the used N$^3$LO results is good  
in the case of the Gross--Llewellyn Smith sum rule (this fact was 
already known from the considerations of Ref.\cite{Samuel}). In the case   
 of $n=2$ and  $n\geq 6$ moments the results are also in 
satisfactory agreement. Indeed, one should keep in mind that the difference 
between the results of column 3 and 4 of Table 2 should be devided by the  
factor $(1/4)^3$, which comes from our definition of expansion parameter 
$A_s=\alpha_s/(4\pi)$. Note, that starting from $n\geq 6$ the results 
of application 
of [0/2] Pad\'e approximants, which in accordance with analysis 
of Ref.\cite{Gardi} are reducing scale-dependence uncertainties, 
are even closer to the the results of the interploation procedure 
(for the comparison of the estimates, given by [1/1] and 
[0/2] Pad\'e approximants in the case of moments of $xF_3$ SF see  
 Ref.\cite{KPS1}, while in Ref.\cite{PP} the similar 
topic  was analysed within the quantum mechanic model).   
For $n=$3,4 the  interpolation method 
gives completely different results. The failure of the application of the 
Pad\'e estimates approach in these cases might be related   to the irregular 
sign structure of the perturbative series under consideration.

\begin{table}
\begin{tabular}{||r|c|c||}
\hline
 &$N_{max}$
 &$\Lambda_{\overline{MS}}^{(4)}$~(MeV)\\
\hline
result of Ref.\cite{KPS1}: NLO  & 6  & 339$\pm$36 \\  
                                & 7  & 340$\pm$37 \\
                                & 8  & 343$\pm$37 \\
                                & 9  & 345$\pm$37 \\
                                &10  & 339$\pm$36 \\
~~~~~~~~~~~~~~~~~~~~~~~NNLO  & 6  & 326$\pm$35 \\ \hline
NNLO results with  &  6   &  325$\pm$35 \\    
inclusion of NNLO & 7 & 326$\pm$31 \\  
terms of $\gamma_{NS,F_3}^{(n)}$ & 8   & 329$\pm$36 \\   
  &   9 & 332$\pm$36  \\  \hline 
N$^3$LO approximate results with          & 6 & 324$\pm$33  \\
inclusion of the interpolated           & 7 & 322$\pm$33  \\
values of $C^{(3)}$(n)-terms 
                   & 8 & 325$\pm$34 \\
           & 9 & 326$\pm$33 \\    
\hline
\end{tabular}
\caption{The results of the fits of the  CCFR data 
for $xF_3$ SF, taking into account the NNLO approximation for 
  $\gamma_{F_3,NS}^{(n)}$. 
The initial scale of RG evolution  is $Q_0^2$=20 GeV$^2$. }
\label{tab:c}
\end{table}

\section{Some results of the fits}

In Table 3  we present the comparison of the results of the determination 
of the 
$\Lambda_{\overline{MS}}^{(4)}$ parameter, made  
in Ref.\cite{KPS1}, with the new ones, obtained by taking into account 
more definite theoretical information. 
Since NNLO corrections to the anomalous 
dimensions and N$^3$LO contributions  to the coefficient 
functions of  odd moments of the  $xF_3$ SF are now known up to 
$n=$13,  it became possible 
to study the dependence of the results of the fits  
from  the value of  $N_{max}$, which we can now vary from $N_{max}=6$ 
to $N_{max}=9$. It should be mentioned that for $N_{max}=6$ the new 
NNLO result and its $Q_0^2$ dependence are in agreement with the 
results of Ref.\cite{KPS1}. However, the incorporation of higher number 
of moments, and thus the  increase  of $N_{max}$,  make the NNLO 
(and  approximate N$^3$LO ) results 
almost independent from the variation of $Q_0^2$ in the interval  
5~GeV$^2$--100~GeV$^2$. This is the welcome feature of including 
into the fits  the results of the  new analytical calculations of the 
NNLO corrections to anomalous dimensions and N$^3$LO corrections to 
the coefficient functions 
of odd moments of the $xF_3$ SF \cite{RV}. 
Comparing now the central values 
of the results 
of the stable  NLO fits of Ref.\cite{KPS1} 
with the new NNLO and N$^3$LO results, 
we observe the decrease of the theoretical uncertainties 
and, probably, the saturation of the predictive power 
of the corresponding perturbative series at the 4-loop level. 
More detailed results  of our fits, including  
extraction of $\alpha_s(M_Z)$, its scale dependence and 
the information about the behaviour of  
 twist-4 corrections at the NNLO and N$^3$LO, in the case of 
$N_{max}=9$, will be described elsewhere \cite{KPS2}.

{\bf Acknowledgements}

We present here some of  the results, reported at Quarks-2000 
International Seminar, Pushkin, May 2000, together 
the first results from Ref.\cite{KPS2}. We wish to thank 
the participants of this productive workshop, and especially 
A. N. Tavkhelidze and F. J. Yndur\'ain for their interest  
and  inspiring discussions. One of us (GP) would 
like to thank the 
Organizing Committee of Quarks-2000 Seminar for 
their hospitality 
in Pushkin and St.Petersburg.

We are also grateful to S. A. Larin for constructive comments on 
the outcome of our previous research \cite{KKPS,KPS1}, summarized in the 
talk at Quarks-2000.

We are grateful to D. V. Shirkov and V. A. Ilyin for presnting  
the results of our previous research in the Plenary Meeting talk
of the ACAT'2000 International Workshop, Fermilab, 16-20 October 2000
and M. Fischler, who   supported the submission of the summary of our 
previous 
results to the Poster Session of ACAT'2000.

It is also a pleasure to thank S. Catani and A. Peterman for discussions of  
subjects related to the material of this  contribution and of  our 
continuing research.

%
%

\end{document}